\documentclass[pra, A4paper, 11pt,amsmath, amsymb, notitlepage]{revtex4-1}

\usepackage{bm,hyperref,tikz,graphicx,calligra,braket}
\usepackage[utf8]{inputenc} 
\newcommand{\bigamma}{\stackrel{\leftrightarrow}{\gamma} }

\begin{document}

\title{Probing the System-Environment Coupling using System Dynamics}
\author{Vinayak Jagadish}
\email[email: ]{vinayak@iisertvm.ac.in}
\affiliation{School of Physics, IISER TVM, CET Campus, Thiruvananthapuram, Kerala, India 695016}

\author{Anil Shaji}
\email[email: ]{shaji@iisertvm.ac.in}
\affiliation{School of Physics, IISER TVM, CET Campus, Thiruvananthapuram, Kerala, India 695016}

\begin{abstract}
Modeling the environment of a single qubit as an $N$ dimensional quantum system, we show that the dynamics of the qubit alone, if measured in sufficient detail, can reveal the parameters of the qubit-environment coupling Hamiltonian. We show that data from quantum process tomography experiments can be used to get information about the environment that can be used to minimize its deleterious effects on the state of the qubit. 
\end{abstract}

\maketitle

Quantum state tomography and quantum process tomography~\cite{chuang97a, dariano01a, emerson07a, howard06a, liu04a, liu05a, liu12a, noguchi11a, obrien04a, schmiegelow11a, yamamoto10a, yuen11a, steffen2006a, noguchi11a} are standard tools in the characterization and development of quantum information processing devices.  In particular, detailed state and process tomography on quantum bits implemented in different types of physical systems are routinely done and the data used to verify the fidelity of initialization procedures, gate operations, readout schemes etc. Tomography of either kind is often necessitated by the inability to completely isolate the quantum system of interest from the uncontrolled influence of other physical systems around it. The effect of the environment leads to deviations in the initialization and dynamics of the quantum systems of interest from the ideal case. Moreover, since the details of the nature and dynamics of the environment are typically unknown, it is not possible to predict what its influence on the system will be. So to get a handle on the unwanted, decohering effects of the environment, direct observation of the states of the system of interest and the processes it undergoes are required.

Can the extensive data obtained about the state and evolution of a open quantum system through tomography be used to gain quantitative information about the environment of the system? This is the main question addressed in this Letter. By extracting as much information as possible about the nature and dynamics of the environment and in particular about the coupling between the system and its environment, it would be possible to identify ways of isolating the system of interest from all unwanted influences of other systems around it. This question was addressed in a very limited context in~\cite{jordan07a} where the system, as well as its environment, are assumed to be single qubits. Here, we remove the restriction that the environment is a single qubit and let it be an arbitrary $N$ level quantum system. Since quite a few experiments involving individual quantum systems have qubits as the system of interest, we let the system in our analysis be a qubit also.  We show how, in principle, the parameters of the system-environment Hamiltonian can be extracted by making sufficiently detailed observations on the system qubit and its dynamics. It is possible to extract partial information about the state of the environment also from the system dynamics but we defer that question to a later time and focus on the coupling Hamiltonian.  We give a detailed example showing the reconstruction of the parameters of the Hamiltonian starting from simulated measurement data of the system qubit.

The state of the system of interest - the qubit - is written in terms of the three Pauli matrices, which are also SU(2), generators, denoted by $\vec{\Sigma} = \big( \Sigma_{1}, \, \Sigma_{2}, \, \Sigma_{3} \big)$.  We often refer to the system qubit as the $\Sigma$-system from here on. The $\Sigma_{i}$ operators satisfy the commutation relations,
\begin{equation}
\label{eq:1}
 [ \Sigma_{i}, \, \Sigma_{j}] = 2i \epsilon_{ijk} \Sigma_{k}. 
 \end{equation}
The environment is assumed to be a general $N$ level quantum system with its state written in terms of the $N^{2}-1$ generators of SU(N) denoted by $\vec{\Lambda}$~\cite{greiner_quantum_1994} (see Appendix A). These generators satisfy the commutation relations, 
\[ [\Lambda_i, \, \Lambda_j]=2if_{ijk}\Lambda_k, \]
where $f_{ijk}$'s are the structure constants of SU(N). The tensor product structure of Hilbert space makes our treatment of the environment ($\Lambda$-system) quite general in that it includes the case where the environment is a collection of lower dimensional quantum systems. The most general Hamiltonian for the $\Sigma$-system interacting with the $N$-level environment, in units where $\hbar = 1$ is
\begin{equation}
\label{eq:hamil1}
	H=\frac{1}{2}\Big(\alpha_j\Sigma_j+\beta_k\Lambda_k+\sum_{j=1}^3\sum_{k=1}^{N^2-1}\gamma_{jk}
\Sigma_j\Lambda_k \Big). 
\end{equation}
 The parameters specifying the Hamiltonian are the three $\alpha$'s, $N^2-1$ $\beta$'s and $3(N^2-1)$ $\gamma$'s. Since $H$ is Hermitian all these parameters have to be real. These are the parameters we are trying to find out by observing the dynamics of the $\Sigma$-system. We use condensed notation in terms of the vectors,
 \begin{eqnarray*}
 	\vec{\alpha} & = & (\alpha_{1}, \, \alpha_{2}, \, \alpha_{3}), \\
	\vec{\beta} & = & (\beta_{1}, \, \ldots, \, \beta_{N^{2}-1} ),
 \end{eqnarray*}
and the tensor 
\begin{equation}
\label{eq:bigamma}
 	\bigamma = \left( \begin{array}{cccc} \gamma_{11} & \gamma_{12} & \ldots & \gamma_{1 N^{2}-1}  \\ \gamma_{21} & \gamma_{22} & \ldots & \gamma_{2 N^{2}-1} \\ \gamma_{31} & \gamma_{32} & \ldots & \gamma_{3 N^{2}-1} \end{array} \right) \equiv \left( \begin{array}{c} \tilde{\gamma_{1}}  \\  \tilde{\gamma_{2}}  \\ \tilde{\gamma_{3}} \end{array} \right),
\end{equation}
so that
\[ H = \frac{1}{2} \big( \vec{\alpha} \cdot \vec{\Sigma} + \vec{\beta} \cdot \vec{\Lambda} + \vec{\Sigma} \cdot \bigamma \cdot \vec{\Lambda} \big). \] 
Note that
\begin{equation}
\label{eq:dot1} 
	\bigamma \cdot \vec{\Lambda}  = \big( \gamma_{1k}\Lambda_{k} , \,  \gamma_{2k} \Lambda_{k} , \, \gamma_{3k} \Lambda_{k}  \big)^{T},
\end{equation}
and
\begin{equation}
\label{eq:dot2}
\vec{\Sigma} \cdot \bigamma  =   (\Sigma_{j}\gamma_{j1}, \, \Sigma_{j}\gamma_{j2}, \, \Sigma_{j}\gamma_{j3}, \, \Sigma_{j}\gamma_{j4}, \, \Sigma_{j}\gamma_{j5}, \, \Sigma_{j}\gamma_{j6}, \, \Sigma_{j}\gamma_{j7}, \, \Sigma_{j}\gamma_{j8}). 
\end{equation}

To obtain the parameters of the Hamiltonian, the $\Sigma$-system is initialised in one of three preparations:
 $\rho^{(1)}_{0} =(1+\Sigma_{1})/2$,  $\rho^{(2)}_{0} = (1+\Sigma_{2})/2$, and $\rho^{(3)}_{0} = (1+\Sigma_{3})/2$. In what follows we assume that we are completely ignorant about the state of the environment and hence ascribe the fully mixed state $\openone_e/N$ to it. If partial information about the state of the environment is available, that can be incorporated into the following analysis in a straightforward manner. In the Schr\"{o}dinger picture, the time evolution of these three states above, generated by $H$, transforms them into
\[ \rho^{(k)}_{t} = \frac{1}{2} \big( 1 + a^{(k)}_{1}(t) \Sigma_{1} + a^{(k)}_{2}(t) \Sigma_{2} + a^{(k)}_{3}(t) \Sigma_{3} \big) , \; k=1,2,3. \]
The nine functions $a^{(k)}_{j}(t)$ are obtained experimentally for some length of time $t$ as part of a complete process tomography experiment. The particular choice of initial test states used here is not the only one that can be made but the computations are simpler for this choice. Any three states whose density operators along with the density matrix of the fully mixed state, span the space of operators on the $\Sigma$-system will suffice. 
Note that $a^{(k)}_{j}(t)$ can be written as expectation values of observables on the time evolved test states as  $a^{(k)}_{j}(t) = \langle \Sigma_{j} \rangle^{(k)}_{t} = {\rm Tr} \big[\rho^{(k)}_{t} \Sigma_{j} \big]$.
Switching to the Heisenberg picture where the time dependence is on the observables and noting that both pictures lead to the same expectation values for observables, we have $a^{(k)}_{j}(t) = \langle \Sigma_{j} (t)\rangle^{(k)} = {\rm Tr} \big[\rho^{(k)}_{0} \Sigma_{j}(t) \big]$. Now consider the $n^{\rm th}$ time derivative of $a^{(k)}_{j}(t)$ in the Heisenberg picture, 
\[ \frac{d^{n} \;}{dt^{n}} a^{(k)}_{j}(t) =  \bigg\langle \frac{d^{n} \;}{dt^{n}}  \Sigma_{j}(t) \bigg\rangle^{(k)} = {\rm Tr} \bigg[ \rho^{(k)}_{0} \frac{d^{n} \;}{dt^{n}} \Sigma_{j}(t) \bigg]. \]
Depending on the time resolution of the experiment that determines $a^{(k)}_{j}(t)$, the higher derivatives of the nine functions that appear on the left hand side of the above equation can be computed to any desired accuracy. 
What remains to be shown is that the time derivatives of the Heisenberg picture Pauli operators that appear on the right hand side are functions of the Hamiltonian parameters that we are trying to find. We also have to show that the time derivatives to increasing order, along with the time derivatives of the experimentally determined $a^{(k)}_{j}(t)$ generate sufficient number of linearly independent equations to solve for all the parameters of $H$.

As an example consider the Heisenberg equation of motion for $\Sigma_{1}$,
\[ \frac{d\;}{dt} \Sigma_{1}(t) \otimes \openone_{e} = i [H, \Sigma_{1}(t) \otimes \openone_{e}]. \]
The Hamiltonian acts on both the qubit and the environment and hence the unit operator on the environment, $\openone_{e}$ is part of the definition of the $\Sigma_{1}$ operator which acts only on the system qubit. However in what follows we will not be carrying the $\openone_{e}$ operator explicitly in the expressions we write. We then have
\begin{eqnarray}
 \dot{\Sigma}_{1}(t) \!\! &= & \! i [H, \, e^{iHt}\Sigma_{1} e^{-iHt}] = ie^{iHt} [ H, \, \Sigma_{1}] e^{-iHt} \nonumber \\
 & =& \! \alpha_2\Sigma_3(t) \! - \! \alpha_3\Sigma_2(t) \! + \! \gamma_{2k}\Lambda_k
\Sigma_3(t) \! -\! \gamma_{3k}\Lambda_k\Sigma_2(t),\quad
 \label{eq:2}
\end{eqnarray}
where we have used Eq.~(\ref{eq:1}) and the Einstein summation convention. For simplicity let us restrict ourselves to small values of $t$ so that $\Sigma_{i}(t) \simeq \Sigma_{i}(0) \equiv \Sigma_{i}$ and $a_{j}^{(k)}(t) \equiv a_{j}^{(k)}$. Using the fact that the trace of a commutator is zero, we have 
\begin{equation}
\label{eq:a13d}
 \dot{a}^{(3)}_{1} = {\rm Tr} \big[ \rho_{0}^{(3)}  \dot{\Sigma}_{1}\big]   =   \frac{1}{2} {\rm Tr}\{ ( \openone + \Sigma_{3}) i[H, \Sigma_{1}] \} = \alpha_{2}.
 \end{equation}
 
The trace in the above equation is over both the $\Sigma$ and $\Lambda$ systems and we have used the fact that the SU(N) generators, $\Lambda_{j}$ are all traceless. Note that in the above expression we have used the short hand, ${\rm Tr} \big[ \rho_{0}^{(3)} \otimes (\openone_{e}/N)  \dot{\Sigma}_{1}(0) \otimes \openone \big] =  {\rm Tr} \big[ \rho_{0}^{(3)}  \dot{\Sigma}_{1}(0)\big] \times 1$.  From here on the unit density matrix of the $\Lambda$-system will be suppressed since it has no bearing on our discussion except that we are admittedly completely ignorant about the state of the environment. 

Note that the restriction to small values of $t$ in the above analysis may be relaxed by redefining $\Sigma_{i} = a_{j}^{(i)}(t) \Sigma_{j}$,  and we can extend the above argument to connect the parameters of the Hamiltonian to the derivatives of the functions $\tilde{a}_{j}^{(k)}(t)$ that are the coefficients of the newly redefined $\Sigma_{i}$ describing the further, short-time, evolution of the three preparations listed earlier. Coming back to the case where $t$ is small, we use the analogues of (\ref{eq:2}) for $\Sigma_{2}$ and $\Sigma_{3}$ as well as the analogues of the traces in Eq.~(\ref{eq:a13d}) to construct the matrix of first order time derivatives of the nine measurable functions $a_{j}^{(k)}$ as 	
\begin{equation}
	\label{eq:firstorder}
	\dot{a} = \left(  \begin{array}{ccc} 0 &  - \alpha_{3} & \alpha_{2} \\  \alpha_{3} & 0 &  -\alpha_{1} \\ -\alpha_{2} & \alpha_{1} & 0\end{array} \right).
\end{equation}
In other words, the first time derivatives of the nine functions $a_{j}^{(k)}$ evaluated at $t=0$ form a real, anti-symmetric, $3\times 3$ matrix whose three independent elements give us three of the parameters of the Hamiltonian, namely $\alpha_{1}$, $\alpha_{2}$ and $\alpha_{3}$. This is not particularly surprising since $\alpha_{j}$'s are the coefficients of the part of the Hamiltonian that act only on the $\Sigma$-system and by examining the dynamics of the $\Sigma$-system one would expect to get information about the $\alpha$'s. In fact if the $\Sigma$-system is assumed to be closed, then the same expressions for the $\alpha$'s follow from the closed dynamics. 

To obtain the remaining parameters, equations connecting higher order time derivatives of $a_{j}^{(k)}$ to functions of these parameters have to be generated. To this end we have to compute higher order commutators of $H$ with the $\Sigma_{i}$'s. To make the expressions compact, we note that for $3 \times 1$ vectors whose components are numbers or vectors like $\vec{\alpha}$ and $\vec{\Sigma}$, the usual cross product is defined in terms of the Levi-Civita tensor $\epsilon_{ijk}$ .\begin{equation} 
\label{eq:cross1}
	[\vec{\alpha} \times \vec{\Sigma} ]_{i} =  \epsilon_{ijk} \alpha_{j} \Sigma_{k}.
\end{equation}
In direct analogy, we define a ``cross product'' for the vector of SU(N) generators and its coefficients as
\begin{equation}
\label{eq:cross2}
 [\vec{\beta} \times \vec{\Lambda}]_{i} \equiv f_{ijk} \beta_{j}\Lambda_{k}. 
\end{equation}
Using Eqs.~(\ref{eq:dot1}), (\ref{eq:dot2}), (\ref{eq:cross1}) and (\ref{eq:cross2}) we can re-write the first time derivative of $\vec{\Sigma}$ in vector notation as,
\begin{equation}
\label{eq:firstcomm}
	\dot{\vec{\Sigma}} = i [H, \, \vec{\Sigma}] = \vec{\alpha} \times \vec{\Sigma} \; + \bigamma \cdot \vec{\Lambda} \times \vec{\Sigma}.
\end{equation}

To generate the equations connecting the observed $\ddot{a}_{j}^{(k)}$ to the parameters of $H$ we have to compute the double commutator, $i[H,\, i[H,\, \vec{\Sigma}]]$ that in the vector notation introduced above is, 
$i[\vec{\alpha} \cdot \vec{\Sigma} + \vec{\beta} \cdot \vec{\Lambda} + \vec{\Sigma} \cdot \bigamma \cdot \vec{\Lambda}, \; \vec{\alpha} \times \vec{\Sigma}  +( \bigamma \cdot \vec{\Lambda}) \times \vec{\Sigma} ]/2$.
The cross products we have introduced are particularly handy in computing the higher order commutations because it reduces essentially to a sequence of replacements of the type $\vec{\Sigma} \rightarrow \vec{X} \times \vec{\Sigma}$ and $\vec{\Lambda} \rightarrow \vec{X} \times \vec{\Lambda}$ and is therefore rather straightforward (see Appendix B for details). Exploiting the computational simplicity provided by our choice of notation, we get, 
\begin{eqnarray*}
	\label{eq:doublec}
	i[H,\, i[H,\, \vec{\Sigma}]] & = & \vec{\alpha} \times \vec{\alpha} \times \vec{\Sigma}  + \vec{\alpha} \times \bigamma \cdot \vec{\Lambda}  \times \vec{\Sigma} +   \bigamma \cdot \vec{\Lambda} \times \vec{\alpha} \times \vec{\Sigma}  \nonumber \\ 
	&& \quad  + \, \bigamma \cdot \vec{\Lambda} \times \bigamma \cdot \vec{\Lambda} \times \vec{\Sigma}  +  \bigamma \cdot(\vec{\beta} \times \vec{\Lambda}) \times \vec{\Sigma} + \bigamma \cdot (\vec{\Sigma} \cdot \bigamma \times \vec{\Lambda} )\times \vec{\Sigma} .
\end{eqnarray*} 
When connecting $\ddot{a}_{j}^{(k)}$ to the double commutator as ${\rm Tr}(\rho_{0}^{(k)} i[H,\,i[H,\, \Sigma_{j}]] )$, the trace operation reduces all terms  containing an odd number of the traceless matrices $\Lambda_{k}$ to zero and we obtain,
\begin{widetext}\begin{equation}
	\label{eq:addot}
	\ddot{a} = -\left( \begin{array}{ccc} \alpha_{2}^{2} + \alpha_{3}^{2} + |\tilde{\gamma}_{2}|^{2} + |\tilde{\gamma}_{3}|^{2} & -\alpha_{1}\alpha_{2} -\tilde{\gamma}_{1} \cdot \tilde{\gamma}_{2} & -\alpha_{1}\alpha_{3} -\tilde{\gamma}_{1} \cdot \tilde{\gamma}_{3} \\ 
	-\alpha_{1}\alpha_{2}-\tilde{\gamma}_{1} \cdot \tilde{\gamma}_{2}  & \alpha_{1}^{2} + \alpha_{3}^{2}+|\tilde{\gamma}_{1}|^{2} + |\tilde{\gamma}_{3}|^{2} & -\alpha_{2}\alpha_{3}-\tilde{\gamma}_{2} \cdot \tilde{\gamma}_{3} 
	\\ -\alpha_{1}\alpha_{3} -\tilde{\gamma}_{1}  \cdot \tilde{\gamma}_{3} & -\alpha_{2} \alpha_{3}-\tilde{\gamma}_{2} \cdot \tilde{\gamma}_{3}  & \alpha_{1}^{2} + \alpha_{2}^{2}+\tilde{\gamma}_{1}^{2} + \tilde{\gamma}_{2}^{2}
	\end{array} \right).
\end{equation}\end{widetext}
With the $\alpha$'s determined from $\dot{a}$, the six independent equations above can be used to find the lengths of the vectors $\tilde{\gamma}_{1}$, $\tilde{\gamma}_{2}$ and $\tilde{\gamma}_{3}$ as well as the dot products $\tilde{\gamma}_{1} \cdot \tilde{\gamma}_{2}$, $\tilde{\gamma}_{1} \cdot \tilde{\gamma}_{3}$ and $\tilde{\gamma}_{2} \cdot \tilde{\gamma}_{3}$. Alternatively we can find up to six of the $3(N^{2}-1)$ parameters $\gamma_{jk}$ by measuring $\ddot{a}$ and solving the equations in (\ref{eq:addot}). 
To generate more equations for solving for the remaining parameters, we go on to the third time derivative of $a_{j}^{(k)}(0)$. We can assemble an expression for the triple commutator that would look quite formidable (see Appendix C) using the replacement rules for cross products mentioned earlier. Noting again that only terms having an even number of $\vec{\Lambda}$s and an odd number of $\vec{\Sigma}$s will not vanish under the trace operation that gives us $\dddot{a}$ we have
\begin{equation}
	\label{eq:tridot}
	\dddot{a}_{j}^{(k)}   =    \epsilon_{jkl}\alpha_{l} \big(|\vec{\alpha}|^{2} + |\tilde{\gamma}_{1}|^{2} + |\tilde{\gamma}_{2}|^{2} + |\tilde{\gamma}_{2}|^{2}\big)   +  2\epsilon_{jkl} \alpha_{m} (\tilde{\gamma}_{m} \cdot \tilde{\gamma}_{l}) + 3 f_{plm} \beta_{p} \gamma_{jl} \gamma_{km}.
\end{equation}
 Eq.~(\ref{eq:tridot}) gives a further set of equations that one can solve either for individual components of the $\gamma$ vectors or for the components of $\vec{\beta}$. The antisymmetry of the both SU(2) and SU(N) structure constants imply that the right hand side of Eq.~(\ref{eq:tridot}) changes sign on exchanging $j$ and $k$. So $\dddot{a}$ is an antisymmetric matrix like $\dot{a}$ and hence we expect to get at most three independent equations for the parameters from the triple time derivative. 
The trace equations with the odd order commutators are antisymmetric and that with the even order commutators are symmetric. Hence we expect to get three independent equations each from the odd orders and six each from the even orders. Assuming an average of 4.5 parameters from each order, we can estimate the minimum order to which commutators are to be computed in order to have sufficient linearly independent equations so as to solve for the $3 + N^{2}-1 + 3(N^{2}-1) = 4 N^{2}-1$ unknown parameters as $(4N^{2}-1)/4.5$. 

The reconstructed values of $\gamma_{ij}$ are not unique since the particular choice of basis in the space of operators on the environment is not expected to have any bearing on the evolution of the qubit. Orthogonal transformations in the space of $\Lambda$ matrices with corresponding rotations of the vectors $\gamma_{1k}$, $\gamma_{2k}$ and $\gamma_{3k}$ that leave $\bigamma \cdot \vec{\Lambda}$ invariant does not have any effect on the observed system dynamics. This freedom in $\gamma_{ij}$ can in fact be used along with a suitable choice of basis in the $\vec{\Sigma}$-space to reduce the number of non-zero entries in the $\gamma_{ij}$ matrix~\cite{jordan07a}.  

In the absence of actual experimental data, to do a numerical example, we start by constructing a Hamiltonian by assigning the following values to the parameters $\alpha_{j}$, $\beta_{j}$ and the $\gamma_{jk}$: $\alpha_1=1$, $\alpha_2=2$, $\alpha_3=3$, $\beta_1=1$, $\beta_2=2$, $\beta_3=1$, $\beta_4=1$, $\beta_5=1$, $\beta_6=1$, $\beta_7=1$, $\beta_8=0.1$, $\gamma_{11}=1$, $\gamma_{22}=1$, $\gamma_{33}=1$. All the rest were set to zero.  Assuming again that the initial state of the environment is fully mixed, we evolved the combined system numerically for the three prototypical initial states for the system qubit namely, $\rho_{0}^{(1)}$, $\rho_{0}^{(2)}$ and $\rho_{0}^{(3)}$. The exact numerical evolution was made to simulate real data to the extent that we computed the reduced density matrix for the evolved system states only at discrete and not too short intervals. This mimics the finite data rate available in the lab. Extra noise could be added to the numerical evolution but for this example no such noise has been added.

With the data, rather artificially discretized, we compute the derivatives $\dot{a_j}^{(k)}$, $\ddot{a_j}^{(k)}$ etc numerically. The computed derivatives are only approximations to the true values of the derivatives because of the finite time steps we have enforced. With real data, the error due to the finite data rate can be mitigated by using the fact that  the matrices of odd and even derivatives of ${a_{j}^{(k)}}$ are antisymmetric and symmetric respectively. Using this knowledge, we make the obtained numerical matrix at each order antisymmetric or symmetric as the case may be.  

\begin{figure}[!htb]
\resizebox{8 cm}{5 cm}{\includegraphics{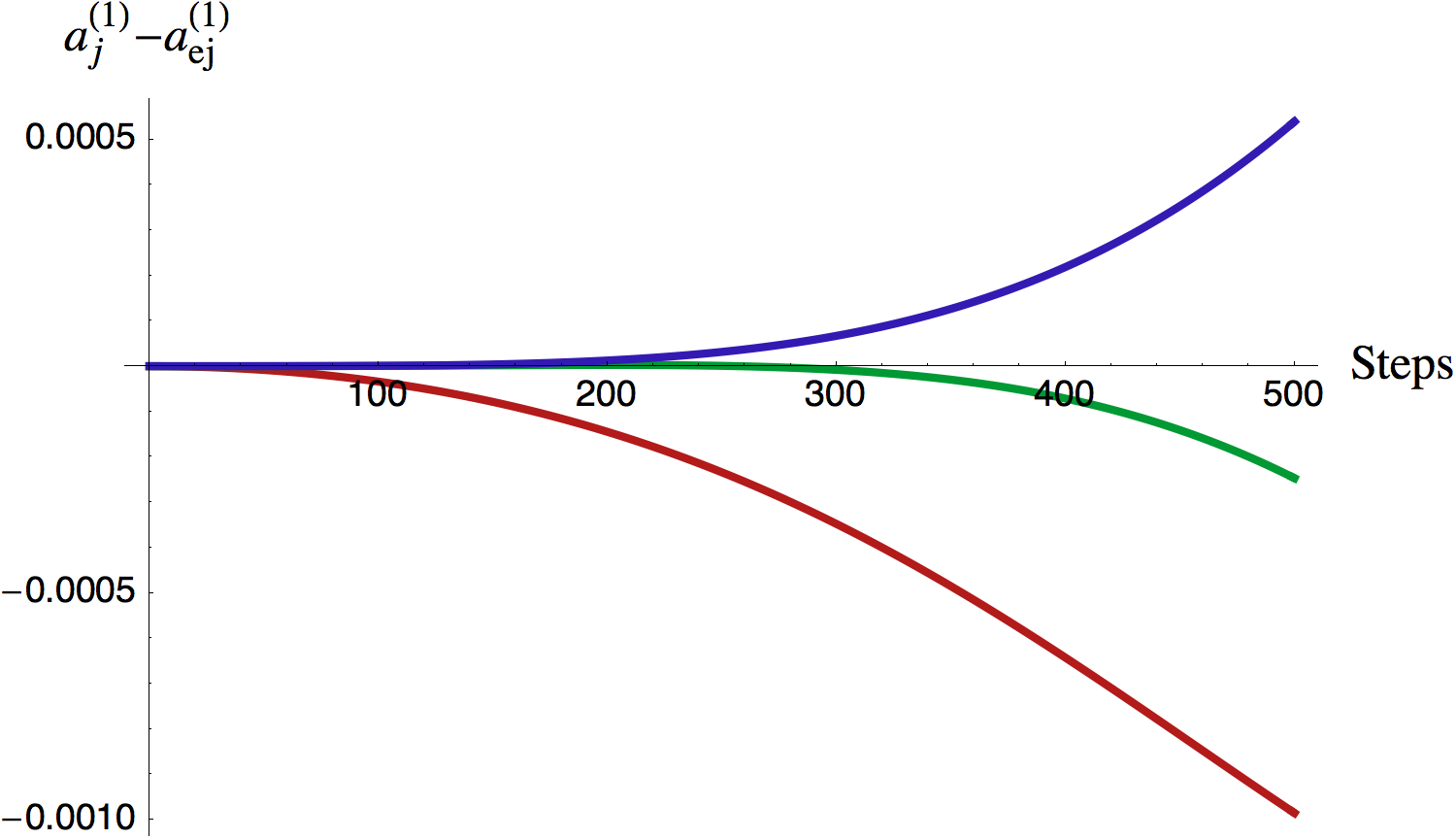}}
\caption{(Color online) The difference between ${a_{j}^{(1)}}$ and ${a_{ej}^{(1)}}$(reconstructed) versus the number of time steps for $j=1,2,3$ are plotted in red, green and blue respectively. The time step used for the simulated data is 0.001 in units where $\hbar=1$. \label{fig1}} 
\end{figure}

To see how good the reconstructed Hamiltonian is, in Fig.~\ref{fig1} we plot the difference between ${a_{j}^{(k)}}$ as computed using the exact Hamiltonian and ${a_{ej}^{(k)}}$ computed using the reconstructed Hamiltonian. We see that the difference even after 500 time steps is only around 0.05 percent of range of values of $a_{j}^{(k)}(t)$ which is between $-1$ and $1$.

To summarise, we have shown that the parameters of the Hamiltonian of a qubit interacting with an $N$ dimensional quantum system can be obtained, in principle, from the time dependance of the qubit alone. The methods presented here can be generalised in a straightforward manner to situations where the system is higher dimensional as well. It is worth noting that the magnitudes of the coupling of the system qubit to the environment given by $|\tilde{\gamma}_{j}|^{2}$ is obtained at the level of $\ddot{a}_{j}^{(k)}$. The dot products $\tilde{\gamma}_{j} \cdot \tilde{\gamma}_{k}$ also obtained from the second derivative tells us about the interdependence, if any, between the environmental influences on the three mutually orthogonal directions of the $\Sigma$ system. This information can in itself be used to minimise the unwanted influence of the environment on a quantum system of interest. It also will aid in identifying decoherence free or decoherence less subspaces as well as in designing improved and robust control pulses and sequences. 

V.~J. thanks CSIR for an SRF. A.~S.~acknowledges the support of the Department of Science and Technology, Government of India, through the Fast-Track Scheme for Young Scientists (SERC Sl.~No.~2786), and the Ramanujan Fellowship programme.
 \appendix
\section{Lie Algebra for SU(N)}
We can construct the $N^{2}-1$ traceless generators $\Lambda_i$, for arbitrary $SU(N)$, using the following method. For every $i,j = 1,2,3,\ldots,N;\; i < j$, we define two $N \times N$ matrices
\[[\Lambda^{\{1\}}(i,j)]_{\mu\nu}=\delta_{j\mu}\delta_{i\nu}+\delta_{j\nu}\delta_{i\mu},\]
\[[\Lambda^{\{2\}}(i,j)]_{\mu \nu}=-i(\delta_{i\mu}\delta_{j\nu}-\delta_{i\nu}\delta_{j\mu}),\]
which form $N(N-1)$ linearly independent matrices. The next $N-1$ matrices are constructed according to
\[\Lambda_{n^2-1}=\sqrt{\frac{2}{n^2-n}} \left(\begin{array}{cccccc}
(1&0&0&0&...&0\\0&1&0&0&...&0\\0&0&1)_{n-1}&0&...&0\\0&0&0&-(n-1)&...&0\\...&...&...&0
&...&...\\0&0&0&0&...&0\end{array}\right)_{N\times N}\]
for $n=2,3,...,N$.
Following this convention, $N^2-1$ traceless matrices can be generated.  These matrices form a representation of the SU(N) generators.

\subsubsection{Example: The SU(3) generators \label{appc}}
 The infinitesimal generators of SU(3) Lie algebra are the Gellmann matrices. The eight linearly independent generators obey the commutation relation:
 \[  [\Lambda_i,\Lambda_j]=2i f^{ijk}\Lambda_k \]
  where the $ f^{ijk}$'s are the completely antisymmetric structure constants.
\[  f^{123}=1, f^{147}= f^{165}= f^{246}= f^{257}= f^{345}= f^{376}=\frac{1}  {2},  f^{458}= f^{678}=\frac{\sqrt3}{2}\]
The generators $\Lambda_{i}$ are given by 
\[  \Lambda_1=\left(\begin{array}{ccc}
0&1&0\\1&0&0\\0&0&0\end{array}\right) \quad  \Lambda_2=\left(\begin{array}{ccc}
0&-i&0\\i&0&0\\0&0&0\end{array}\right) \quad  \Lambda_3=\left(\begin{array}{ccc}
1&0&0\\0&-1&0\\0&0&0\end{array}\right)\]
\[ \quad \Lambda_4= \left(\begin{array}{ccc}
0&0&1\\0&0&0\\1&0&0\end{array}\right)\quad \Lambda_5=\left(\begin{array}{ccc}
0&0&-i\\0&0&0\\i&0&0\end{array}\right)\quad \Lambda_6=\left(\begin{array}{ccc}
0&0&0\\0&0&1\\0&1&0\end{array}\right)\] 
\[ \Lambda_7= \left(\begin{array}{ccc}
0&0&0\\0&0&-i\\0&i&0\end{array}\right) \quad \Lambda_8=\frac{1}{\sqrt3} \left(\begin{array}{ccc}
1&0&0\\0&1&0\\0&0&-2\end{array}\right)\]
There are three independent SU(2) subgroups:\{$\Lambda_1$,$\Lambda_2$,x)\},\{$\Lambda_4$,$\Lambda_5$,y\}and \{$\Lambda_6$,$\Lambda_7$,z\} where x,y, z are linear combinations of $\Lambda_3$ and $\Lambda_8$.

 \section{The double commutator \label{appA}}
 The double commutator we have to evaluate is
 \[ i[H,\, i[H,\, \vec{\Sigma}]] = \frac{i}{2}[\vec{\alpha} \cdot \vec{\Sigma} + \vec{\beta} \cdot \vec{\Lambda} + \vec{\Sigma} \cdot \bigamma \cdot \vec{\Lambda}, \; \vec{\alpha} \times \vec{\Sigma}  +( \bigamma \cdot \vec{\Lambda}) \times \vec{\Sigma} ]. \]
The individual terms in the commutator are
\begin{eqnarray*}
	[\vec{\alpha} \cdot \vec{\Sigma}, \, \vec{\alpha} \times \vec{\Sigma}  ]_{j} & = & \alpha_{k} \epsilon_{jlm} \alpha_{l} [\Sigma_{k}, \Sigma_{m}] = 2i \epsilon_{jlm} \epsilon_{pkm} \alpha_{k} \alpha_{l} \Sigma_{p} \\
	& = & 2 i \epsilon_{jlm} \alpha_{l} (-\epsilon_{mkp} \alpha_{k} \Sigma_{p}) = -2i  \epsilon_{jlm} \alpha_{l} [\vec{\alpha} \times \vec{\Sigma}]_{m} \\
	& = & -2i [\vec{\alpha} \times \vec{\alpha} \times \vec{\Sigma}]_{j}.
\end{eqnarray*}
The subscript $j$ on both sides of the equation denotes a component of the vector equation. Putting the components back into a vector equality, we have
\[ [\vec{\alpha} \cdot \vec{\Sigma}, \, \vec{\alpha} \times \vec{\Sigma}  ] = -2i \, \vec{\alpha} \times \vec{\alpha} \times \vec{\Sigma}. \]
\[ [\vec{\beta} \cdot \vec{\Lambda}, \,  \vec{\alpha} \times \vec{\Sigma}] = 0.\]
\begin{eqnarray*}
[ \vec{\Sigma} \cdot \bigamma \cdot \vec{\Lambda}, \, \vec{\alpha} \times \vec{\Sigma}]_{j} & = &  (\bigamma \cdot \vec{\Lambda})_{k} \epsilon_{jlm} \alpha_{l} [\Sigma_{k}, \Sigma_{m}] = 2i \epsilon_{jlm} \epsilon_{pkm} (\bigamma \cdot \vec{\Lambda})_{k}  \alpha_{l} \Sigma_{p} \\
	& = & 2 i \epsilon_{jlm} \alpha_{l} \{ -\epsilon_{mkp} (\bigamma \cdot \vec{\Lambda})_{k}  \Sigma_{p} \} = -2i  \epsilon_{jlm} \alpha_{l} [(\bigamma \cdot \vec{\Lambda})\times \vec{\Sigma}]_{m} \\
	& = & -2i [\vec{\alpha} \times \bigamma \cdot \vec{\Lambda}  \times \vec{\Sigma}]_{j},
\end{eqnarray*}
and so we have
\[ [ \vec{\Sigma} \cdot \bigamma \cdot \vec{\Lambda}, \, \vec{\alpha} \times \vec{\Sigma}] = -2i \, \vec{\alpha} \times \bigamma \cdot \vec{\Lambda}  \times \vec{\Sigma}. \]
From the above three results we see a pattern emerging. We have
\begin{equation}
\label{eq:pattern} 
	[\vec{X} \cdot \vec{\Sigma}, \, \vec{Y} \times \vec{\Sigma}  ] = -2i \, Y \times X \times \vec{\Sigma}, 
\end{equation}
where $\vec{X}$ and $\vec{Y}$ are arbitrary vectors with numerical or operator components which are not functions of the $\Sigma$'s. Using Eq.~(\ref{eq:pattern}) we have 
\[ [\vec{\alpha} \cdot \vec{\Sigma}, \, \bigamma \cdot \vec{\Lambda} \times \vec{\Sigma}] = -2i\bigamma \cdot \vec{\Lambda} \times \vec{\alpha} \times \vec{\Sigma}. \]
Now we look at the two terms in the double commutator that contain commutations between the SU(N) generators, 
\begin{eqnarray*} 
	[ \vec{\beta} \cdot \Lambda, \, \bigamma \cdot \vec{\Lambda} \times \vec{\Sigma}] ]_{j} & = & \epsilon_{jlm} \beta_{k} \gamma_{lp} \Sigma_{m} [\Lambda_{k}, \, \Lambda_{p}] =2i\, \epsilon_{jlm} \beta_{k} \gamma_{lp} \Sigma_{m} f_{qkp} \Lambda_{q} \\
	& = &2i\, \epsilon_{jlm} \gamma_{lp} (-f_{pkq} \beta_{k} \Lambda_{q}) \Sigma_{m} = -2i \, \epsilon_{jlm} \gamma_{lp} [\vec{\beta} \times \vec{\Lambda}]_{p} \Sigma_{m} \\
	& = & -2i \, \epsilon_{jlm} [\bigamma \cdot(\vec{\beta} \times \vec{\Lambda})]_{l} \Sigma_{m} \\
	& = & -2i [\bigamma \cdot(\vec{\beta} \times \vec{\Lambda}) \times \vec{\Sigma}]_{j}.
\end{eqnarray*}
 So we have
 \[ [ \vec{\beta} \cdot \Lambda, \, \bigamma \cdot \vec{\Lambda} \times \vec{\Sigma}] ] = -2i \, \bigamma \cdot(\vec{\beta} \times \vec{\Lambda}) \times \vec{\Sigma}. \]
 Finally
 \begin{eqnarray*}
 	[\vec{\Sigma} \cdot \bigamma \cdot \vec{\Lambda}, \, \bigamma \cdot \vec{\Lambda} \times \vec{\Sigma}]_{j} & = & \gamma_{kl} \epsilon_{jmn} \gamma_{mp} [\Sigma_{k} \Lambda_{l}, \, \Sigma_{n} \Lambda_{p}] \\
	& = & 2i \, \epsilon_{jmn} \epsilon_{qkn} \gamma_{kl} \gamma_{mp} \Sigma_{q} \Lambda_{l} \Lambda_{p} + 2i \, \epsilon_{jmn} f_{qlp} \gamma_{kl} \gamma_{mp} \Sigma_{k} \Sigma_{n} \Lambda_{q} \\
	& =& 2i \, \epsilon_{jmn} (\bigamma \cdot \vec{\Lambda})_{m} \{-\epsilon_{nkq} (\bigamma \cdot \vec{\Lambda})_{k} \Sigma_{q} \} + 2i \,  \epsilon_{jmn} \gamma_{mp} \{ -f_{plq} (\vec{\Sigma} \cdot \bigamma )_{l} \Lambda_{q}  \}  \Sigma_{n}\\
	& = & -2i \, \epsilon_{jmn} (\bigamma \cdot \vec{\Lambda})_{m} (\bigamma \cdot \vec{\Lambda} \times \vec{\Sigma})_{n} - 2i \,  \epsilon_{jmn} \gamma_{mp} \{ (\vec{\Sigma} \cdot \bigamma) \times \vec{\Lambda} \}_{p} \Sigma_{n} \\
	& = &  -2i \, [(\bigamma \cdot \vec{\Lambda})\times (\bigamma \cdot \vec{\Lambda}) \times \vec{\Sigma}]_{j} - 2i \, [\bigamma \cdot \{(\vec{\Sigma} \cdot \bigamma) \times \vec{\Lambda} \} \times \vec{\Sigma}]_{j},
 \end{eqnarray*}
 As a vector equation, 
\[ [\vec{\Sigma} \cdot \bigamma \cdot \vec{\Lambda}, \, \bigamma \cdot \vec{\Lambda} \times \vec{\Sigma}] = -2i \, (\bigamma \cdot \vec{\Lambda}) \times (\bigamma \cdot \vec{\Lambda}) \times \vec{\Sigma} - 2i \, [\bigamma \cdot \{(\vec{\Sigma} \cdot \bigamma) \times \vec{\Lambda} \} ]\times \vec{\Sigma}\]
Looking at commutators involving both $\Sigma$'s and $\Lambda$'s the pattern that emerges is 
\begin{equation}
	\label{eq:rule1} 
	[\vec{X} \cdot  \vec{\Sigma}, \, f(\vec{\Sigma})] = -2i \, f(\vec{X} \times \vec{\Sigma}), 
\end{equation}
and
\begin{equation}
	\label{eq:rule2}
	[\vec{X} \cdot \vec{\Lambda}, \, g(\vec{\Lambda})] = -2i \, g(\vec{X} \times \vec{\Lambda}), 
\end{equation}
where $f$ and $g$ are arbitrary linear functions of the operator valued vectors $\vec{\Sigma}$ and $\vec{\Lambda}$ respectively involving the dot and cross products of these vectors. We also have a ``product rule'',
\begin{equation}
	\label{eq:rule3}
	[\vec{\Sigma} \cdot \bigamma \cdot \vec{\Lambda}, \, h(\Sigma, \Lambda)] = -2i \, h(\bigamma \cdot \vec{\Lambda} \times \Sigma, \Lambda) - 2i \, h(\Sigma, \vec{\Sigma} \cdot \bigamma \times \Lambda), 
\end{equation}
for an arbitrary bilinear function $h$ of $\vec{\Sigma}$ and $\vec{\Lambda}$. Putting all the previous results in this section together, we obtain
\begin{eqnarray}
	\label{eq:doublec}
	i[H,\, i[H,\, \vec{\Sigma}]] & = & \frac{i}{2} \big\{ -2i \, \vec{\alpha} \times \vec{\alpha} \times \vec{\Sigma}  -2i \, \vec{\alpha} \times \bigamma \cdot \vec{\Lambda}  \times \vec{\Sigma} -2i \, \bigamma \cdot(\vec{\beta} \times \vec{\Lambda}) \times \vec{\Sigma} -2i\bigamma \cdot \vec{\Lambda} \times \vec{\alpha} \times \vec{\Sigma}\nonumber \\
	&& \qquad - \; 2i \, \bigamma \cdot \vec{\Lambda} \times \bigamma \cdot \vec{\Lambda} \times \vec{\Sigma} - 2i \, \bigamma \cdot (\vec{\Sigma} \cdot \bigamma \times \vec{\Lambda} )\times \vec{\Sigma} \big\} \nonumber \\
	& = & \vec{\alpha} \times \vec{\alpha} \times \vec{\Sigma}  + \vec{\alpha} \times \bigamma \cdot \vec{\Lambda}  \times \vec{\Sigma}  +  \bigamma \cdot \vec{\Lambda} \times \vec{\alpha} \times \vec{\Sigma} + \bigamma \cdot \vec{\Lambda} \times \bigamma \cdot \vec{\Lambda} \times \vec{\Sigma}\nonumber \\
	&& \qquad  + \;   \bigamma \cdot(\vec{\beta} \times \vec{\Lambda}) \times \vec{\Sigma} + \bigamma \cdot (\vec{\Sigma} \cdot \bigamma \times \vec{\Lambda} )\times \vec{\Sigma} .
\end{eqnarray} 

\section{The triple commutator \label{appB}}

Using the double commutator from Eq.~(5) and the rules for computing the commutators involving each of the three terms in $H$ as given in Eqs.~(2), (3) and (4), we can compute the different pieces of the triple commutator $i[H,\, i[H,\, i[H,\, \vec{\Sigma}]]]$ as follows
\begin{eqnarray}
	\label{eq:tterm1}
	\frac{i}{2}[\vec{\alpha} \cdot \vec{\Sigma} + \vec{\beta} \cdot \vec{\Lambda} + \vec{\Sigma} \cdot \bigamma \cdot \vec{\Lambda} , \, \vec{\alpha} \times \vec{\alpha} \times \vec{\Sigma} ] & = & \vec{\alpha} \times \vec{\alpha} \times \vec{\alpha} \times \vec{\Sigma} + \vec{\alpha} \times \vec{\alpha} \times \bigamma \cdot \vec{\Lambda} \times \vec{\Sigma}.
\end{eqnarray}
\begin{eqnarray}
	\label{eq:tterm2}
	\frac{i}{2}[\vec{\alpha} \cdot \vec{\Sigma} + \vec{\beta} \cdot \vec{\Lambda} + \vec{\Sigma} \cdot \bigamma \cdot \vec{\Lambda} , \, \vec{\alpha} \times \bigamma \cdot \vec{\Lambda}  \times \vec{\Sigma}] & = & \vec{\alpha} \times \bigamma \cdot \vec{\Lambda} \times \vec{\alpha} \times \vec{\Sigma} + \vec{\alpha} \times \bigamma \cdot (\vec{\beta} \times \vec{\Lambda})\times \vec{\Sigma} \nonumber \\
	&& \vec{\alpha} \times \bigamma \cdot (\vec{\Sigma} \cdot \bigamma \times \vec{\Lambda}) \times \vec{\Sigma} +  \vec{\alpha} \times \bigamma \cdot \vec{\Lambda} \times \bigamma \cdot \vec{\Lambda} \times \vec{\Sigma}. \quad
\end{eqnarray}
\begin{eqnarray}
	\label{eq:tterm3}
	\frac{i}{2}[\vec{\alpha} \cdot \vec{\Sigma} + \vec{\beta} \cdot \vec{\Lambda} + \vec{\Sigma} \cdot \bigamma \cdot \vec{\Lambda} , \,  \bigamma \cdot \vec{\Lambda}  \times \vec{\alpha} \times \vec{\Sigma}] & = &  \bigamma \cdot \vec{\Lambda} \times \vec{\alpha} \times \vec{\alpha} \times \vec{\Sigma} +  \bigamma \cdot (\vec{\beta} \times \vec{\Lambda})\times\vec{\alpha} \times \vec{\Sigma} \nonumber \\
	&&  \bigamma \cdot (\vec{\Sigma} \cdot \bigamma \times \vec{\Lambda}) \times\vec{\alpha} \times \vec{\Sigma} +   \bigamma \cdot \vec{\Lambda} \times \vec{\alpha} \times \bigamma \cdot \vec{\Lambda} \times \vec{\Sigma}. \quad
\end{eqnarray}
\begin{eqnarray}
	\label{eq:tterm4}
	\frac{i}{2}[\vec{\alpha} \cdot \vec{\Sigma} + \vec{\beta} \cdot \vec{\Lambda} + \vec{\Sigma} \cdot \bigamma \cdot \vec{\Lambda} , \,  \bigamma \! \cdot \vec{\Lambda}  \times \bigamma \! \cdot \vec{\Lambda} \times \vec{\Sigma}] & = &  \bigamma \cdot \vec{\Lambda} \times \bigamma \cdot \vec{\Lambda} \times \vec{\alpha} \times \vec{\Sigma} +  \bigamma \cdot (\vec{\beta} \times \vec{\Lambda})\times \bigamma \cdot \vec{\Lambda} \times \vec{\Sigma} \nonumber \\
	&& +  \; \bigamma \! \cdot \vec{\Lambda} \times \bigamma \! \cdot (\vec{\beta} \times \vec{\Lambda})\times \vec{\Sigma} + \bigamma \! \cdot (\vec{\Sigma} \cdot \bigamma \!\times \vec{\Lambda}) \times \bigamma \! \cdot \vec{\Lambda} \times \vec{\Sigma} \nonumber \\
	&& + \;  \bigamma \! \cdot \vec{\Lambda} \times \bigamma \cdot (\vec{\Sigma} \cdot \bigamma \times \vec{\Lambda})  \times \vec{\Sigma}  \nonumber \\
	&& + \;  \bigamma \cdot \vec{\Lambda} \times \bigamma \cdot \vec{\Lambda} \times \bigamma \cdot \vec{\Lambda} \times \vec{\Sigma}. \quad
\end{eqnarray}
\begin{eqnarray}
	\label{eq:tterm5}
	\frac{i}{2}[\vec{\alpha} \cdot \vec{\Sigma} + \vec{\beta} \cdot \vec{\Lambda} + \vec{\Sigma} \cdot \bigamma \cdot \vec{\Lambda} , \,   \bigamma \cdot(\vec{\beta} \times \vec{\Lambda}) \times \vec{\Sigma}] & = &   \bigamma \cdot(\vec{\beta} \times \vec{\alpha} \times \vec{\Lambda}) \times \vec{\Sigma} +   \bigamma \cdot(\vec{\beta} \times \vec{\beta} \times \vec{\Lambda}) \times \vec{\Sigma} \nonumber \\
	&&  \bigamma \cdot(\vec{\beta} \times \vec{\Sigma} \cdot \bigamma \times \vec{\Lambda}) \times \vec{\Sigma} +    \bigamma \cdot(\vec{\beta} \times \vec{\Lambda}) \times \bigamma \cdot \vec{\Lambda} \times \vec{\Sigma}. \qquad 
\end{eqnarray}
\begin{eqnarray}
	\label{eq:tterm6}
	\frac{i}{2}[\vec{\alpha} \cdot \vec{\Sigma} + \vec{\beta} \cdot \vec{\Lambda} + \vec{\Sigma} \cdot \bigamma \cdot \vec{\Lambda} , \,   \bigamma \cdot (\vec{\Sigma} \cdot \bigamma \times \vec{\Lambda} )\times \vec{\Sigma} ] & = &    \bigamma \cdot [(\vec{\alpha} \times \vec{\Sigma}) \cdot \bigamma \times \vec{\Lambda}]\times \vec{\Sigma} +   \bigamma \cdot (\vec{\Sigma} \cdot \bigamma \times \vec{\beta} \times \vec{\Lambda} )\times \vec{\Sigma} \nonumber \\
	&&  \bigamma \! \cdot (\vec{\Sigma} \cdot \bigamma \! \times  \vec{\Lambda} )\times \vec{\alpha} \times \vec{\Sigma} +   \bigamma \! \cdot (\vec{\Sigma} \cdot \bigamma \! \times \vec{\Sigma} \cdot \bigamma \times \vec{\Lambda} )\times \vec{\Sigma}  \nonumber \\
	&& \bigamma \cdot [(\bigamma \cdot \vec{\Lambda} \times \vec{\Sigma}) \cdot \bigamma \times \vec{\Lambda} ]\times \vec{\Sigma} \nonumber \\
	&& \bigamma \cdot (\vec{\Sigma} \cdot \bigamma \times \vec{\Lambda} )\times \bigamma \cdot \vec{\Lambda} \times \vec{\Sigma}
\end{eqnarray}
\bibliography{arxiv}

 \end{document}